\documentstyle{mn}

\def\Wg{$\Omega_g$}
\def\nhi{\mbox{$N_{\rm HI}$}}

\def\q0{q$_0$}
\def\lya{Ly$\alpha$\ }

\def\etal{\rm et al.}
\def\eg{\protect\rm e.g.}
\def\ie{\protect\rm i.e.}

\def\simgt{$_>\atop{^\sim}$}

\def\hi{\mbox{\rm HI}}

\def\apj{ApJ }

\title[Evolution of Neutral Gas at High Redshift]
{Evolution of Neutral Gas at High Redshift -- \\
Implications for the Epoch of Galaxy Formation} 
\author[Storrie-Lombardi, McMahon \& Irwin]
{L. J. Storrie-Lombardi$^{1,\star}$
R. G. McMahon$^1$, 
\& M. J. Irwin$^2$ \\
$^1$Institute of Astronomy, Madingley Road, Cambridge CB3 0HA \\
$^2$Royal Greenwich Observatory, Madingley Road, Cambridge CB3 0EZ \\
$^\star$ current address: 
Observatories of the Carnegie Institution of
Washington, 813 Santa Barbara Street, Pasadena, CA 91101 USA \\
email:  lisa@ociw.edu (LJSL); rgm@ast.cam.ac.uk (RGM); mike@ast.cam.ac.uk (MJI)}

\begin{document}
\maketitle


\begin{abstract}

Though observationally rare, damped \lya absorption systems 
dominate the mass density of neutral gas in the Universe.  
Eleven high redshift damped \lya systems covering
2.8 $<$ z $<$ 4.4 were discovered in 26 QSOs from 
the APM z$>$4 QSO Survey, 
extending these absorption system surveys to the highest redshifts 
currently possible. Combining our new data set with previous surveys
we find that the cosmological mass density in neutral gas,
\Wg, does not rise as steeply prior to z$\sim$2 
as indicated by previous studies. 
There is evidence in the observed \Wg\ for a flattening
at z$\sim$2 and a possible turnover at z$\sim$3.
When combined with the decline at z $>$ 3.5 in number density
per unit redshift of damped systems with column densities
log \nhi$\ge21$ atoms cm$^{-2}$, these results point to an epoch at z \simgt\ 3
prior to which the highest column density damped systems
are still forming. 
We find that over the redshift range 2 $<$ z $<$ 4 the total 
mass in neutral gas is
marginally comparable with the total visible mass in stars in present
day galaxies. However, if one considers the total mass visible in
stellar disks alone, \ie\ excluding galactic bulges, the two values are
comparable. We are observing a mass of neutral gas comparable 
to the mass of visible disk stars.
Lanzetta, Wolfe \& Turnshek  found that
$\Omega(z \approx 3.5)$ was twice $\Omega(z \approx 2)$,
implying a much larger amount of star formation must
have taken place between z$=$3.5 and z$=$2 than is indicated
by metallicity studies.  This created a `cosmic G-dwarf problem'.
The more gradual evolution of \Wg\ we find
alleviates this.  These results
have profound implications for theories of galaxy formation.

\end{abstract}


\section{Introduction}
While the baryonic content of spiral galaxies
that are observed in the present epoch is concentrated in stars,
in the past this must have been in the form of gas.  
The principal gaseous
component in spiral galaxies is HI which has led
to surveys for absorption systems detected by the damped lines  
they produce (Wolfe \etal\ 1986 [WTSC]; Lanzetta \etal\ 1991 [LWTLMH]; 
Lanzetta, Wolfe, \& Turnshek 1995 [LWT], Wolfe \etal\ 1995).
Damped \lya absorption systems have neutral hydrogen column 
densities of \nhi$> 2 \times 10^{20}$ atoms cm$^{-2}$ and  
they dominate the baryonic mass contributed by HI.  
We extend the earlier work on damped \lya systems 
to higher redshifts using twenty-six QSOs from the 
APM Damped \lya Survey (Storrie-Lombardi \etal\ 1996 [SMIH], 
Storrie-Lombardi, Irwin \& McMahon 1996 [SIM]),
with eleven candidate or confirmed damped Ly$\alpha$ 
absorption systems 
covering the redshift range $2.8\le z \le 4.4$ (8 with z $>$ 3.5).
These data more than triple the redshift path surveyed at z $>$ 3
and allow the first systematic study up to z $=$ 4.7.

\section{Evolution of $\Omega_g$ -- Baryons in Neutral Gas} 

The mean cosmological mass density contributed by \lya
absorbers can be estimated as 
\begin{equation}
\langle \Omega_g \rangle = {H_0 \mu m_H \over c \rho_{crit}} 
\int_{N_{min}}^\infty Nf(N,z)dN
\label{eqnomega}
\end{equation}
giving the current mass density in units of the current 
critical density (LWTLMH). 
$\mu$ is the mean molecular weight of the gas which
is taken to be 1.3 (75\% H and 25\% He by mass), 
$m_{\rm H}$ is the mass of the hydrogen atom, 
$\rho_{\rm crit}$ is the current critical mass density, 
$N_{\rm min}$ is the low end of the \hi\ column density
range being investigated,
and $f(N,z)$ is the column density distribution function.  
Unfortunately $f(N,z)$ is not a simple function and its 
evolution with redshift is difficult to accurately quantify 
(LWT, SIM). 
The integral in equation~\ref{eqnomega} can be estimated using 
\begin{equation}
\int_{N_{min}}^\infty Nf(N,z)dN = {\sum_i N_i(\hi) \over \Delta X},
\label{eqnestfn}
\end{equation}
where $\Delta X$ is the absorption distance interval.  
The absorption distance $X$ is
used to remove the redshift dependence in the sample
and put everything on a comoving coordinate scale. 
If the population of absorbers is nonevolving (\ie\ the product
of their space density multiplied by their cross-section does not
change with redshift) they have a constant number
density per unit absorption distance. In a standard 
Friedmann universe $X$ is defined as
\begin{equation}
X(z) = \cases {{2 \over 3}[(1+z)^{3/2} - 1] &if q$_0=0.5$;\cr
              {1 \over 2}[(1+z)^2 - 1]    &if q$_0=0$.}
\label{absdis2eqn}
\end{equation}
(Bahcall \& Peebles 1969; cf.~Tytler 1987).
The errors in \Wg\ are
also difficult to estimate without knowing $f(N,z)$. 
LWTLMH used the standard error in the distribution of \nhi\
which yields zero error if all the column densities in a bin are 
the same.  We have estimated the fractional variance in \Wg\ 
by comparing the observed distribution of $f(N,z)$ with the
equivalent Poisson sampling process.  This gives 
\begin{equation}
\Bigl( {\Delta\Omega_g \over \Omega_g} \Bigr )^2 =
\sum_{i=1}^p N_i^2 / \Bigl(\sum_{i=1}^p N_i\Bigr )^2  
\label{eqnerr}
\end{equation}
and $1/\sqrt{p}$ fractional errors if all the 
column densities included in a bin are equal.
To address uncertainties in $f(N,z)$ 
we also calculated a maximum likelihood estimate of
the errors in the HI column density. We used 
the power law with an exponential turnover
form of the column density distribution function,
\ie\ the gamma-distribution from SIM
\begin{equation}
f(N,z)=( f_* / N_* ) ( N / N_* )^{-\beta} e^{-N/N_*},
\label{gamdisteqn}
\end{equation}
with log $N_*=21.63\pm0.35$,
$\beta=1.48\pm0.30$, and $f_*=1.77\times 10^{-2}$.
Unlike a pure power law this form has a finite integral mass.
The maximum-likelihood estimates of the errors 
agree well with the fractional variance.

LWT found that \Wg\ 
inferred from studies of damped systems rises with 
increasing redshift for 0.008 $<$ z $<$ 3.5. 
For the range 3.0 $<$ z $<$ 3.5 which included 
4 damped systems and was the  
the highest redshift bin in the study, 
$\Omega(z \approx 3.5)$ was twice $\Omega(z \approx 2)$.
This implied a much larger amount of star formation must
have taken place between z$=$3.5 and z$=$2 than is indicated
by metallicity studies.
Specifically, Pettini \etal\ (1994) measured a low mean metallicity
in damped \lya systems at z$\approx$2.2 inferring that
they are observed prior to the bulk of star formation in the disk.
The LWT result also implied 
that the bulk of stars in nearby 
galaxies should be metal poor whereas only a small fraction 
of disk stars in the solar neighbourhood are metal poor.
This result presented  
a `cosmic G-dwarf problem' similar to the 
`G-dwarf problem' described in Schmidt (1963) 
that comes about if bright and faint 
stars formed at the same rate and there was no accretion
on to the disk of the Galaxy. 
They concluded that 
the characteristic epoch of metal production in galaxies
occurred after the characteristic epoch of star formation
and that damped \lya absorbers might trace disk as well as 
spheroid evolution.
Fall \& Pei (1993; and Pei \& Fall 1995) have argued 
that obscuration caused by dust in damped \lya systems could lead 
to significant underestimates of the neutral gas fraction 
at all redshifts, particularly in the range 1 $\le$ z $\le$ 2,
also possibly explaining these results. 

To further investigate these issues, the high redshift confirmed 
and candidate damped \lya systems discovered in the 
APM Damped \lya Survey
(SMIH, SIM) have been combined with lower redshift samples 
(WTSC, LWTLMH, LWT)
to study the evolution of \Wg\ over the range 
0.008 $<$ z $<$ 4.7. The combined data set includes 44 
absorbers in 366 quasars. 
The results are tabulated in table 1 and 
shown in figure 1(a) for q$_0$=0
and 1(b) q$_0$=0.5 (H$_0$=50) following the format presented 
in Wolfe \etal\ (1995). 
The solid bins include the entire data set and the dashed bins 
exclude the new high redshift APM data\footnote{Excluding the new APM data 
effectively yields the data set analyzed in LWT.}.
The region $\Omega_{star}$ is the $\pm1\sigma$ range for the 
mass density in stars in nearby galaxies
(Gnedin \& Ostriker 1992). The point at z$=$0 is 
the value inferred from 21 cm emission from local galaxies 
(Fall \& Pei 1993; Rao \& Briggs 1993).
The most striking result is \Wg\ does not rise 
as steeply prior to z$\sim$2 as indicated by previous studies.  
There is now evidence for a flattening in \Wg\ at z$\sim$2 
and a possible turnover at z$\sim$3.
This result, combined with the decline at z$>$3.5 in number density 
per unit redshift of damped systems with column densities
log \nhi$\ge21$ atoms cm$^{-2}$ (SIM) points to an epoch at z \simgt\ 3
prior to which the largest damped systems are still forming. 
The decrease in number density at high redshift of the highest column
density absorbers is in marked contrast to the more numerous
lower column density systems, \eg\
Lyman-limit systems (\nhi $\sim 10^{18}$ atoms cm$^{-2}$) 
(Storrie-Lombardi \etal\ 1994; Stengler-Larrea \etal\ 1995)
and \lya forest absorbers 
(\nhi $\sim 10^{13}$--$10^{15}$  cm$^{-2}$) \cite{Williger94}.

The inclusion of the APM survey data for z $>$ 3 reduces significantly 
the value previously found for \Wg\ in the bin 3 $<$ z $<$ 3.5.
Only one damped system is added to the
existing four in this redshift range, but the absorption
distance is doubled, which comes in to the calculation 
of \Wg\ in the denominator in equation~\ref{eqnestfn}
(see table 2). The additional redshift path 
added by the APM survey is shown graphically by the 
sensitivity function in figure 6 of SMIH. 
We find that over the redshift range 2 $<$ z $<$ 4 the total mass in
neutral gas 
(\Wg) is marginally comparable with the total visible 
mass in stars in present
day galaxies ($\Omega_{star}$) for q$_0$=0.5. 
However, if one considers the 
total mass visible in stellar disks alone, \ie\ excluding galactic bulges, 
the two values are comparable. Using the result from 
Schechter \& Dressler (1987) that 
galactic disks and bulges contribute equally to the mass density 
of the Universe, we are observing a mass of 
neutral gas comparable to the mass of visible disk stars, \ie\ \Wg $\sim$ 
$\Omega_{disk-stars}$. We note that the uncertainty in the total
mass in visible stars in the local Universe is comparable with
our estimates of the mass in neutral gas at z $>$ 2.
Given this, and the fact that we do not know if damped systems
are the precursors to galactic disks, bulges, or both,  
these results are difficult to interpret.
If we make a plausible correction for obscuration by dust 
as advocated by Pei \& Fall (1995) the \Wg\ points shown
in figure 1(b) would migrate to the positions of the open circles 
shown in figure 2.  
More work is needed to determine the 
severity of dust obscuration in optically selected QSO surveys.
An estimated 20\% correction for the 
neutral gas not in damped \lya systems is shown by the open squares. 

\section{Implications for Galaxy Formation Theories}

The shape of the \Wg\ curve has been used by numerous authors
to constrain theories of galaxy formation and cosmological models
\cite{Klypin95,KauffmannCharlot94,MaBert94,MoMiralde94}.
They have found that cold$+$hot dark matter (CHDM) models are incompatible 
with the previous results from the damped \lya systems at 
z$\sim$3 as they predict too few galactic halos. 
These models need to be reevaluated now that the value of 
\Wg(z$=$3) indicated from our new observations has changed 
significantly. For example,
\cite{Klypin95} show a peak in \Wg\ at z$=$3 for various CHDM models. 
Though the error bars on \Wg\ are still far too large to 
accurately discriminate between details of cosmological models,
the overall shape has implications for structure formation. 

\section{Summary}

Combining the new data from the APM high redshift survey
extends studies of the cosmological mass density in neutral 
gas, \Wg, to z$=$4.7.
We find evidence for a flattening in \Wg\ at 
z$\sim$2 and a possible turnover at z$\sim$3.
Though the turnover is not formally significant,
when combined with the decline at z $>$ 3.5 in number density 
of damped systems with column densities
log \nhi$\ge21$ atoms cm$^{-2}$, these results point to an epoch 
prior to which the largest damped systems are still forming. 

Previous studies indicated that
$\Omega(z \approx 3.5)$ was twice $\Omega(z \approx 2)$,
implying a much larger amount of star formation must
have taken place between z$=$3.5 and z$=$2 than is indicated
by metallicity studies. The more gradual evolution we find
in \Wg\ alleviates this problem.  
The results are consistent with observations of 
damped \lya systems at z$\approx$2.2 that show
a low mean metallicity,   
suggesting that they are observed prior to the bulk 
of star formation in the disk.
We have also made an estimated correction to \Wg\ to 
account for bias in optically selected quasar
samples due to possible obscuration by dust 
in foreground absorbers.
Theories of galaxy formation and constraints on cosmological models 
utilising \Wg\ should be reevaluated in light of 
the new observational results presented here.  
The error bars are still very large at high redshift and to 
differentiate between a peak versus a flattening in the 
\Wg\ curve, larger samples of bright z $>$ 3.5 quasars are needed
to discover damped \lya systems with z $>$ 3.

\bigskip
\noindent{\bf Acknowledgments}

We acknowledge fruitful discussions with Art Wolfe, Max Pettini and
Mike Fall. RGM thanks the Royal Society for support. 
LSL acknowledges support from an Isaac Newton Studentship, the Cambridge
Overseas Trust, and a University of California President's Postdoctoral
Fellowship.  We thank the PATT for time awarded to do 
the observations with the William Herschel Telescope that made
this work possible. We thank the referee, Ken Lanzetta, for his comments.

\clearpage

\begin{table}
\caption{Data for Figures}
\label{t_figdata}
\begin{tabular}{rrrrrrrr}
Bin \ \ \ \ \ \ DLA& &\multicolumn{1}{c}{\# of \ }&\multicolumn{2}{c}{q$_0=0$}
&\multicolumn{2}{c}{q$_0=0.5$}\\
\multicolumn{1}{l}{Redshift \ \  $\langle{\rm z}\rangle$}&
\multicolumn{2}{r}{$\Delta$z \ \ DLA QSO}&
$\Delta X$&\multicolumn{1}{c}{$\Omega_g$}& $\Delta X$&\multicolumn{1}{c}{$\Omega_g$}\\
\multicolumn{1}{l}{Range}&&\multicolumn{1}{c}{in bin}& &[$\times h_{50} 10^{3}$]
& &[$\times h_{50} 10^{3}$]\\
\hline
.008-1.5 \ \ 0.64&47.8&4 \ \ 186&73.1 &0.56$\pm$0.32&58.5&0.70$\pm$0.40\\
1.5-2.0 \ \ 1.89&27.9&4 \ \ 126&79.5 &1.21$\pm$0.71&47.1&2.05$\pm$1.19\\
2.0-3.0 \ \ 2.40&120.2&22 \ \ 176&415.9&1.50$\pm$0.49&223.4&2.79$\pm$0.91\\
3.0-3.5 \ \ 3.17&24.3&5 \ \ \ \ 82&102.0&1.48$\pm$0.72&49.8&3.04$\pm$1.48\\
3.5-4.7 \ \ 4.01&19.2&9 \ \ \ \ 32&93.8 &0.85$\pm$0.34&42.5&1.87$\pm$0.75\\
\multicolumn{7}{l}{Dashed bins, excluding high redshift data.} \\
2.0-3.0 \ \ 2.38&114.6&21 \ \ 154&394.8&1.56$\pm$0.52&212.6&2.90$\pm$0.96\\
3.0-3.5 \ \ 3.19& 11.8&4  \ \ \ \ 56& 48.9&2.79$\pm$1.47& 24.0&5.68$\pm$3.00\\
\hline
\end{tabular}
\raggedright
\end{table}

\begin{table}
\caption{Redshift Path and Absorption Distance}
\label{t_zpath}
\begin{tabular}{lrrr}
\hline
Data Set& $\Delta z$ & $\Delta X$ & $\Delta X$\\
 & & ($q_0=0$) & ($q_0=0.5$)  \\
\hline
{\bf $3.0 < z < 3.5$} & & & \\
APM Damped \lya Survey  &  12.5 &  53.1 &  25.8  \\
WTSC $+$ LWTLMH $+$ LWT &  11.8 &  48.9 &  24.0  \\
Combined                &  24.3 & 102.0 &  49.8  \\
 & & & \\
{\bf $z > 3$} & & & \\
APM Damped \lya Survey  &  30.5 & 141.2 &  65.5  \\
WTSC $+$ LWTLMH $+$ LWT &  13.0 &  54.6 &  26.7  \\
Combined                &  43.5 & 195.8 &  92.2  \\
 & & & \\
{\bf $0 < z < 4.7$} & & & \\
APM Damped \lya Survey  &  36.1 & 162.3 &  76.5  \\
WTSC $+$ LWTLMH $+$ LWT & 203.4 & 602.1 & 344.8  \\
Combined                & 239.5 & 764.4 & 421.3  \\
\hline
\end{tabular}
\end{table}


\begin{figure}
\caption
{The mean cosmological mass density
in neutral gas, \Wg,
contributed by damped \lya absorbers for 0.008$\le$z$\le$4.7
for (a) q$_0$=0 and (b) q$_0$=0.5.
The solid bins include the combined data set and
the dashed bins exclude the new APM high redshift data.
The region $\Omega_{star}$ is the $\pm1\sigma$ range for the
mass density in stars in nearby galaxies
(Gnedin \& Ostriker 1992).  The point at z$=$0 is
the value inferred from 21 cm emission from local galaxies
(Fall \& Pei 1993; Rao \& Briggs 1993).  These results
are tabulated in table 1.}
\label{f_omega}
\end{figure}


\begin{figure}
\caption
{The mean cosmological mass density in
neutral gas, \Wg,
contributed by damped \lya absorbers for 0.008$\le$z$\le$4.7.
The solid bins are the combined data set shown in
figure 1(b) for q$_0=0.5$.  The circles show the observed data
points corrected for possible dust obscuration using values
determined from the closed-box/outflow models shown in
figure 4(b) of Pei \& Fall (1995).  
The squares add
an estimated 20\% correction for neutral gas 
not in damped \lya absorbers.
The region $\Omega_{star}$ is the $\pm1\sigma$ range for the
mass density in stars in nearby galaxies
(Gnedin \& Ostriker 1992).  The point at z$=$0 is
the value inferred from 21 cm emission from local galaxies
(Fall \& Pei 1993; Rao \& Briggs 1993).}
\label{f_dust}
\end{figure}

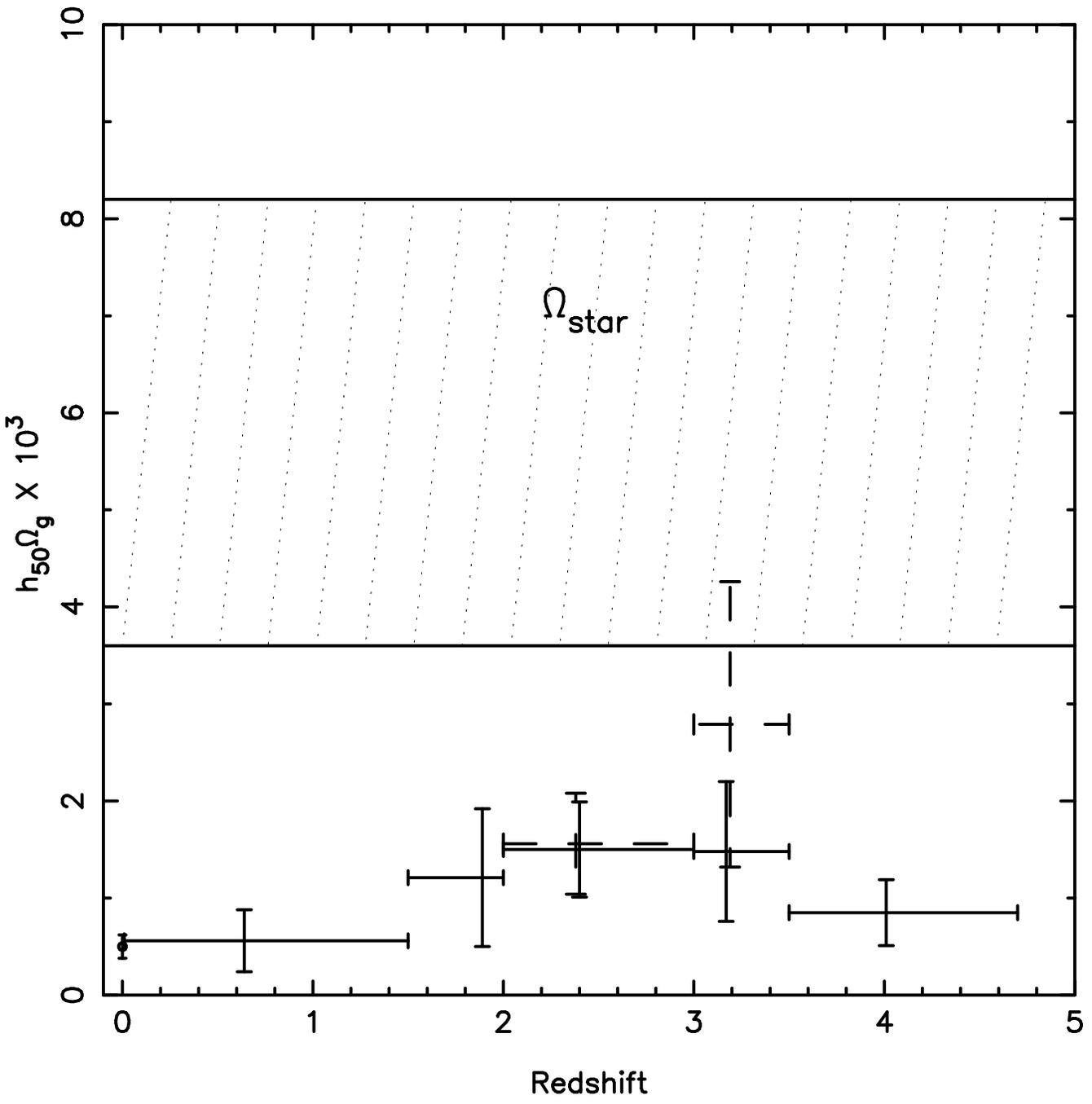
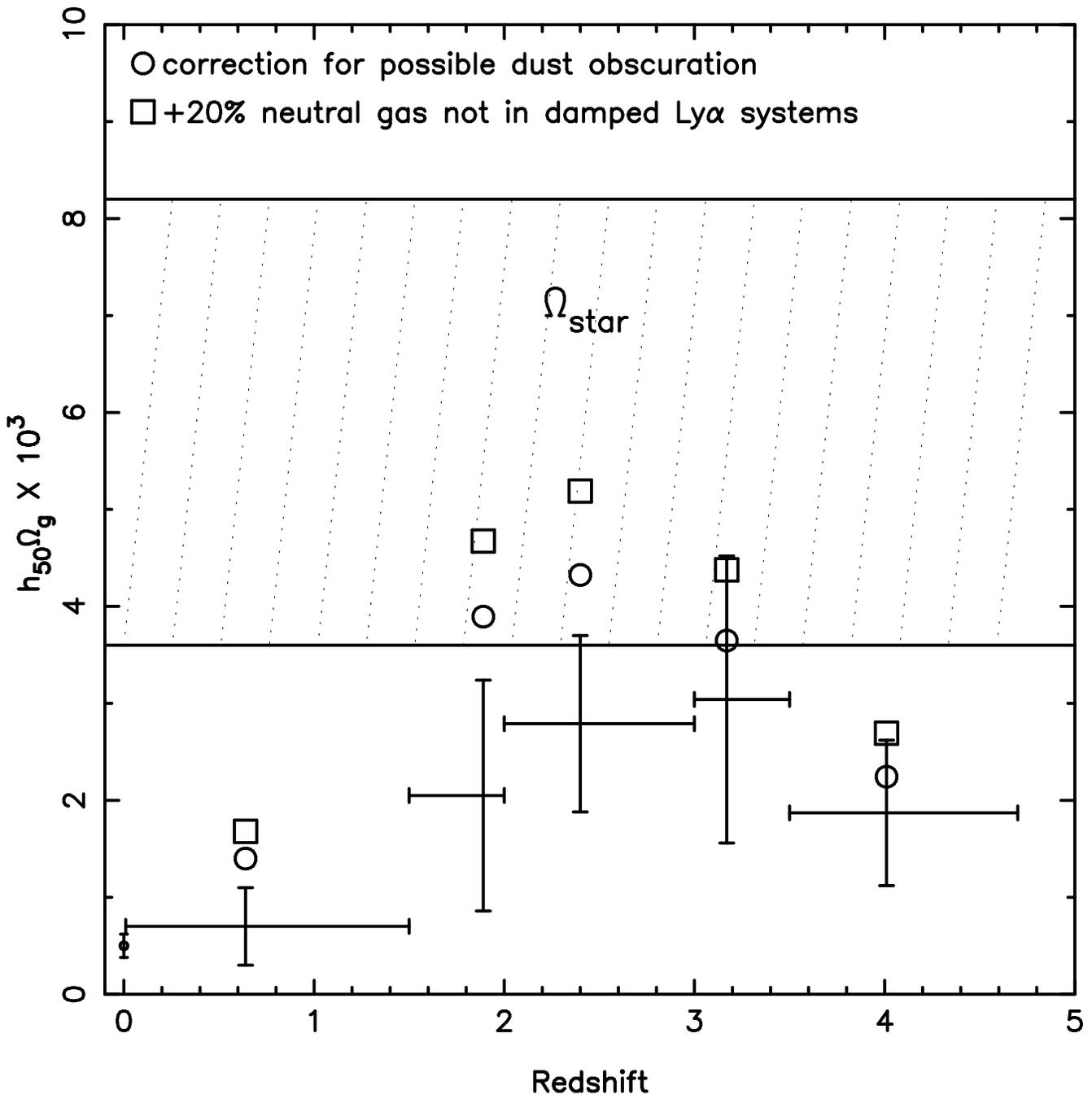
\end{document}